\documentclass[twocolumn,aps,showpacs,prb]{revtex4-1}

\usepackage{graphicx}
\usepackage{epstopdf}
\usepackage{amsmath}
\usepackage{multirow}

\begin{document}

\title{First-principles study on the double-dome superconductivities in kagome material CsV$_3$Sb$_5$ under pressure}

\author{Jian-Feng Zhang}
\author{Kai Liu}\email{kliu@ruc.edu.cn}
\author{Zhong-Yi Lu}\email{zlu@ruc.edu.cn}

\affiliation{Department of Physics and Beijing Key Laboratory of Opto-electronic Functional Materials $\&$ Micro-nano Devices, Renmin University of China, Beijing 100872, China}

\date{\today}

\begin{abstract}

Recent high pressure experiments discovered abnormal double-dome superconductivities in the newly-synthesized kagome materials $A$V$_3$Sb$_5$ ($A$ = K, Rb, Cs), which also host abundant emergent quantum phenomena such as charge density wave (CDW), anomalous Hall effect, nontrivial topological property, etc. In this work, by using first-principles electronic structure calculations, we have studied the CDW state, superconductivity, and topological property in CsV$_3$Sb$_5$ under pressures ($<$ 50 GPa). Based on the electron-phonon coupling theory, our calculated superconducting $T_\text{c}$s are consistent with the observed ones in the second superconducting dome at high pressure, but are much higher than the measured values at low pressure. The further calculations including the Hubbard U indicate that with modest electron-electron correlation the magnetism on the V atoms exists at low pressure and diminishes gradually at high pressure. We thus propose that the experimentally observed superconductivity in CsV$_3$Sb$_5$ at ambient/low pressures may still belong to the conventional Bardeen-Cooper-Schrieffer (BCS) type but is partially suppressed by the V magnetism, while the superconductivity under high pressure is fully conventional without invoking the magnetism. We also predict that there are a second weak CDW state and topological phase transitions in CsV$_3$Sb$_5$ under pressures. Our theoretical assertion calls for future experimental examination.

\end{abstract}

\pacs{}

\maketitle

\section{INTRODUCTION}

A new family of kagome materials $A$V$_3$Sb$_5$ ($A$ = K, Rb, Cs)~\cite{1stprm} has attracted intensive attentions recently~\cite{z2prl,yan,yangsy,sc,sc2,sc3,sc4,sc5,pdw,cdwsc,cdwsc2,cdwsc3,cdw,cdw2,cdw3,cdw4,cdw5,cdw6,cdw7,cdw8,c2cdw,c2cdw2,c2cdw3,3dcdw,c2,cdwtopo,cdwtopo2,z2prl,ahc,ahc2,mag,mag2,press1,press2,press3,lowpress1,lowpress2,lowpress3,Kpress,fengstm} due to their abundant emergent quantum phenomena such as superconductivity~\cite{z2prl,yan,yangsy,sc,sc2,sc3,sc4,sc5,cdwsc,cdwsc2,cdwsc3,pdw,fengstm}, charge density wave (CDW)~\cite{3dcdw,cdw,cdw2,cdw3,c2cdw2,cdw4,cdw5,cdw6,cdw7,cdw8,cdwtopo,cdwtopo2,cdwsc2,cdwsc,cdwsc3,c2cdw,c2cdw3}, nontrivial topological property~\cite{z2prl,cdwtopo,cdwtopo2}, anomalous Hall effect~\cite{ahc,ahc2}, and so on.
With the decreasing temperature, these kagome materials firstly undergo a 2$\times$2$\times$2 CDW transition~\cite{3dcdw,cdw,cdw2,cdw3,c2cdw2,cdw4,cdw5,cdw6,cdw7,cdw8,cdwtopo,cdwtopo2,cdwsc2,cdwsc3,cdwsc,c2cdw,c2cdw3}, then show another unidirectional charge order with the reduction of rotation symmetry from C$_6$ to C$_2$~\cite{c2cdw,c2cdw2,c2cdw3,c2}, and finally enter the superconducting region ($T_\text{c}=$ 0.9$\sim$2.5 K)~\cite{z2prl,yan,yangsy,sc,sc2,sc3,sc4,sc5,cdwsc,cdwsc2,cdwsc3,pdw,fengstm}.
Regarding the characteristics of the superconductivity, different experimental approaches gave distinct results. The thermal conductivity measurement ~\cite{lowpress2} suggests a nodal superconducting gap, but the penetration depth~\cite{sc2} and nuclear magnetic resonance~\cite{mag} experiments propose a nodeless gap. Meanwhile, the scanning tunneling microscopy (STM) measurement~\cite{fengstm} observes the coexistence of nodal and nodeless superconducting gaps.
From the theoretical side, the density functional theory (DFT) calculation with the deformation potential approximation indicates that the conventional electron-phonon coupling (EPC) cannot afford the observed superconducting $T_\text{c}$~\cite{yan}. The DFT plus dynamical mean-field theory (DMFT) calculations also suggest that the local correlation strength in these materials is too weak to generate unconventional superconductivity~\cite{yangsy}. With the absence of long-range magnetic order but the possible existence of magnetic fluctuations or orbital order~\cite{1stprm,mag,mag2}, the superconducting mechanism in these kagome materials is still under debate~\cite{mag,yan,yangsy,sc,sc2,cdwsc,cdwsc2,cdwsc3,pdw,fengstm}.

Beyond the measurements at ambient pressure, recent high-pressure experiments have reported that this family of kagome materials displays similar superconducting behavior under pressure without the structural phase transition~\cite{press1,press2,press3,lowpress1,lowpress2,lowpress3,Kpress}. The pressure can firstly suppress the CDW state and meanwhile enhance the superconductivity. After the optimal pressure point, the superconducting $T_\text{c}$ starts to decrease and then raises again at higher pressures, forming a second superconducting dome. Among these kagome materials, CsV$_3$Sb$_5$ owns the highest superconducting $T_\text{c}$ and its $T_\text{c}$ completely drops to 0 K around 12 GPa, demonstrating distinct double superconducting domes. The pressure evolution of the superconductivities in these kagome materials needs urgent theoretical elucidation, which may help us understand their superconducting mechanism as well as the relationship among the CDW state, superconductivity, and possible magnetism in the geometrically frustrated materials.

In this work, taking CsV$_3$Sb$_5$ as a prototype of these kagome materials, we have performed first-principles calculations on the phonon spectra and the EPC strength to study the double-dome superconducting behavior under pressure. Unexpectedly, in the pressure range we investigated ($<$ 50 GPa), the calculated superconducting $T_\text{c}$ always keeps nonzero, which is much higher than the observed values at low pressure but agrees well with the measured ones at high pressure. The further DFT+U calculations indicate that the modest electron correlation can induce the magnetism on the V atoms while the pressure can effectively suppress the local magnetic moments. Combining these results, we deduce that it is the magnetism (much likely ferrimagnetism) that restrains the conventional superconducting $T_\text{c}$ at ambient/low pressure, forming the first superconducting dome. Our calculations also predict a new weak CDW phase and the topological phase transitions under high pressures, which needs future experimental examination.

\section{Method}

The electronic structure, topological property, and phonon spectra of CsV$_3$Sb$_5$ under pressure were investigated based on the density functional theory (DFT)~\cite{dft1, dft2} and density functional perturbation theory (DFPT)~\cite{dfpt,dfptreview} calculations as implemented in the Quantum ESPRESSO (QE) package~\cite{pwscf}. The interactions between electrons and nuclei were described by the norm-conserving pseudopotentials~\cite{ncpp}. The valence electron configurations were $5s^2 5p^6 6s^1$ for Cs, $3s^2 3p^6 3d^3 4s^2$ for V, and $4d^{10} 5s^2 5p^3$ for Sb, respectively. The generalized gradient approximation (GGA) of Perdew-Burke-Ernzerhof (PBE)~\cite{PBE} type was adopted for the exchange-correlation functional. The kinetic energy cutoff of plane-wave basis was set to be 80 Ry. A 12$\times$12$\times$8 {\bf k}-point mesh was used for the Brillouin zone (BZ) sampling. The Gaussian smearing method with a width of 0.004 Ry was employed for the Fermi surface broadening. The lattice constants were fixed at the experimental values of Ref.~\onlinecite{press1} and the internal atomic positions were fully relaxed until the forces on all atoms were smaller than 0.0002 Ry/Bohr. The spin-orbit coupling (SOC) was included in the band structure and topological property calculations.

The superconductivity in CsV$_3$Sb$_5$ was studied based on the electron-phonon coupling (EPC) theory with the Electron-Phonon Wannier (EPW) software\cite{epw}, which is interfaced with the QE~\cite{pwscf} and Wannier90~\cite{mlwf} packages. The 6$\times$6$\times$4 {\bf k}-mesh and 6$\times$6$\times$2 {\bf q}-mesh were adopted as coarse grids and the 48$\times$48$\times$32 {\bf k}-mesh and 24$\times$24$\times$16 {\bf q}-mesh were utilized as dense grids. The superconducting transition temperature $T_c$ was calculated with the McMillan-Allen-Dynes formula~\cite{mcmillan1, mcmillan2},
\begin{equation}
{T_c=\frac{\omega_{log}}{1.2}\text{exp}[\frac{-1.04(1+\lambda)}{\lambda(1-0.62\mu^*)-\mu^*}]},
\end{equation}
where $\lambda=\sum_{{\bf q}\nu}\lambda_{{\bf q}\nu}$ is the total EPC strength and $\omega_{log}$ is the logarithmic average of the Eliashberg spectral function $\alpha^2F(\omega)$~\cite{Eliashberg},
\begin{equation}
\omega_{log}=\text{exp}[\frac{2}{\lambda}\int{\frac{d\omega}{\omega}\alpha^2F(\omega){ln}(\omega)}],
\end{equation}
\begin{equation}
\alpha^2F(\omega)=\frac{1}{2}\sum_{{\bf q}\nu}\delta(\omega-\omega_{{\bf q}\nu})\omega_{{\bf q}\nu}\lambda_{{\bf q}\nu},
\end{equation}
here $\omega_{{\bf q}\nu}$ is the phonon frequency of the $\nu$-th mode at the wave vector {\bf q} and $\mu^*$ is the effective screened Coulomb repulsion constant, which was empirically set to 0.05 and 0.10 for comparison~\cite{mustar1,mustar2}.

To study the possible magnetism of the V atoms and the correlation effect among the V-3$d$ electrons, we carried out the GGA+U calculations~\cite{ldau} by employing the Vienna Ab-initio Simulation Package (VASP)~\cite{vasp1,vasp2}. Several effective Hubbard U values were adopted in order to check the influence of electron correlation strength on the magnetism.

\section{Results and Discussion}


\begin{figure*}[tb]
\includegraphics[angle=0,scale=0.73]{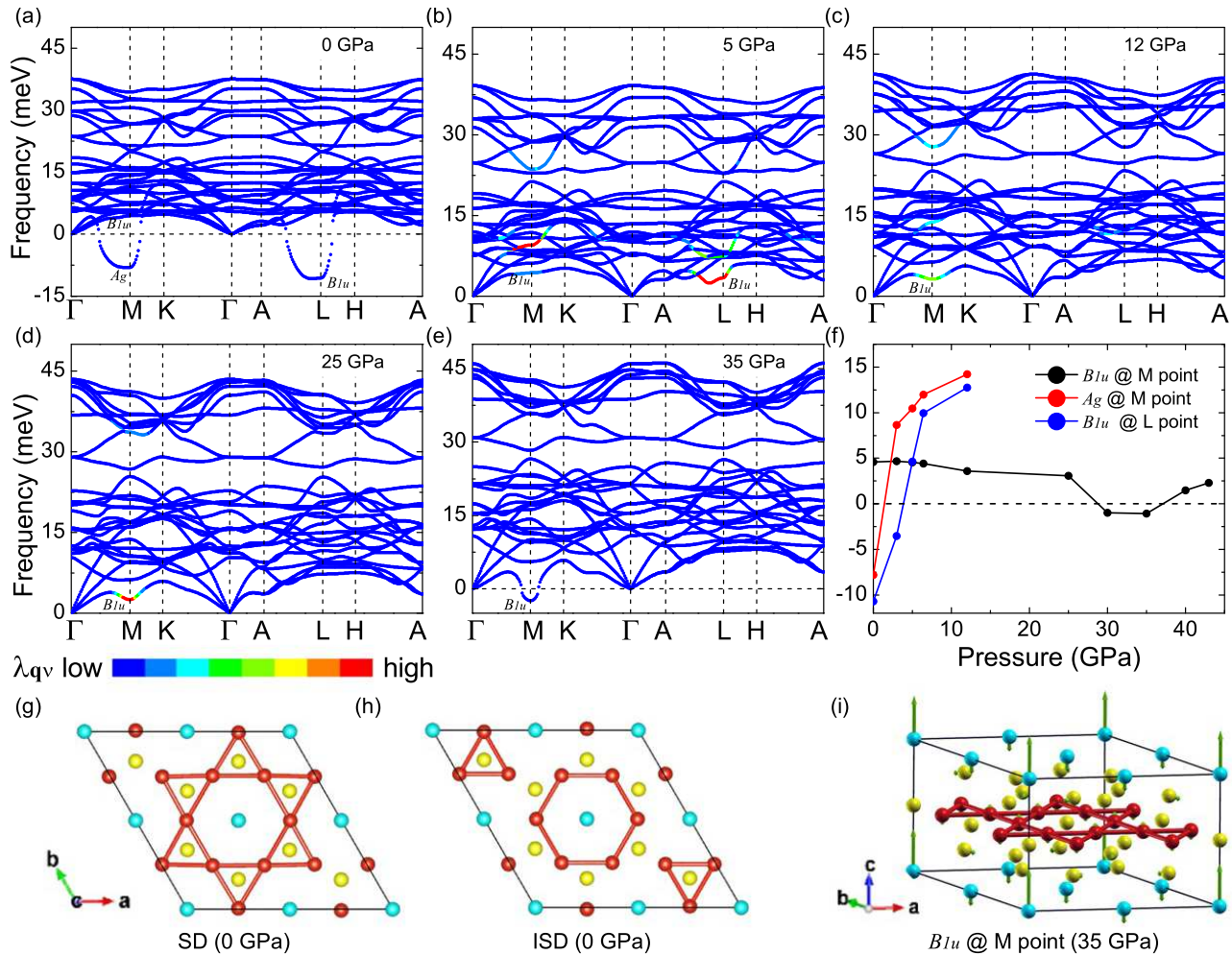}
\caption{(Color online) Phonon spectra of CsV$_3$Sb$_5$ at (a) 0, (b) 5, (c) 12, (d) 25, and (e) 35 GPa, respectively. The line colors in the panels (b)-(d) correspond to the moment- and mode- resolved electron-phonon coupling strength $\lambda_{\bf q\nu}$. The $B_{1u}$ and $A_g$ modes at the $M$ point and the $B_{1u}$ mode at the $L$ point are labeled. (f) Pressure evolution of the phonon frequencies of the $B_{1u}$ (black squares) and $A_g$ (red dots) modes at the $M$ point and the $B_{1u}$ mode (blue triangles) at the $L$ point. (g) and (h) Charge density wave (CDW) distortions for the imaginary $A_g$ mode at the $M$ point under ambient pressure, so called the (g) SD and (h) ISD patterns. The blue, red, and yellow balls represent the Cs, V, and Sb atoms, respectively. The distortion of the $B_{1u}$ mode at the $L$ point is equivalent to a layer by layer stacking of the panels (g) and (h). (i) Distortion for the imaginary $B_{1u}$ mode at the $M$ point under 35 GPa, which is mainly contributed by the movements of Cs atoms along the $c$ axis.}
\label{fig2}
\end{figure*}


We firstly studied the pressure-suppressed CDW state in CsV$_3$Sb$_5$. Five experimental representative pressure conditions were picked out: 0 GPa (ambient pressure), 5 GPa (the pressure at which the CDW state is suppressed and the first superconducting dome appears), 12 GPa (the pressure at which the first superconducting dome is completely suppressed), and 25/35 GPa (the pressures in the second superconducting dome). Figures 1(a)-1(e) respectively show the calculated phonon spectra of CsV$_3$Sb$_5$ under the above pressures. At 0 GPa [Fig. 1(a)], the phonon spectra exhibit a large portion of imaginary frequencies around the $M$ point ($A_g$ mode) and the $L$ point ($B_{1u}$ mode), which can induce the experimentally observed 2$\times$2$\times$2 CDW state~\cite{3dcdw}. The corresponding structure distortions are the theoretically proposed patterns of star of David [SD, Fig. 1(g)] and inverse star of David [ISD, Fig. 1(h)]~\cite{yan}. At 5 GPa, the imaginary phonon modes around the $M$ and $L$ points vanish, indicating that the pressure can suppress the CDW state, which agrees with the previous experimental study~\cite{press1}.
With further increasing pressure, a $B_{1u}$ mode at the $M$ point tends to perform an abnormal softening and eventually transfers to imaginary frequency under high pressure [Figs. 1(c)-1(e)]. In Fig. 1(f), we summarize the pressure evolution of the frequencies for the $B_{1u}$ mode (black dots) and the $A_g$ mode (red dots) at the $M$ point as well as the $B_{1u}$ mode at the $L$ point (blue dots). Apparently, the CDW state (CDW I) is completely suppressed around 4 GPa, while a new 2$\times$2$\times$1 CDW state (CDW II) around 29$\sim$37 GPa is induced, as indicated by the imaginary frequency of the $B_{1u}$ mode at the $M$ point. The structural distortion of this CDW II state is displayed in Fig. 1(i), which mainly consists of the movements of Cs atoms along the $c$ axis. This is quite different from the SD and ISD patterns in the CDW I state, which involve the distortions of V atoms in the $ab$ plane. It's worth noting that although the distortion of the Cs atoms in the CDW II state can exceed 0.1 \AA, the corresponding energy gain is very small. Meanwhile, the Cs distortion in the CDW II state has a slight influence on the electronic states, as shown in the Supplementary Materials~\cite{sm}. Thus, the CDW II state can not effectively affect the superconductivity at high pressure.

\begin{figure}[tb]
\includegraphics[angle=0,scale=0.57]{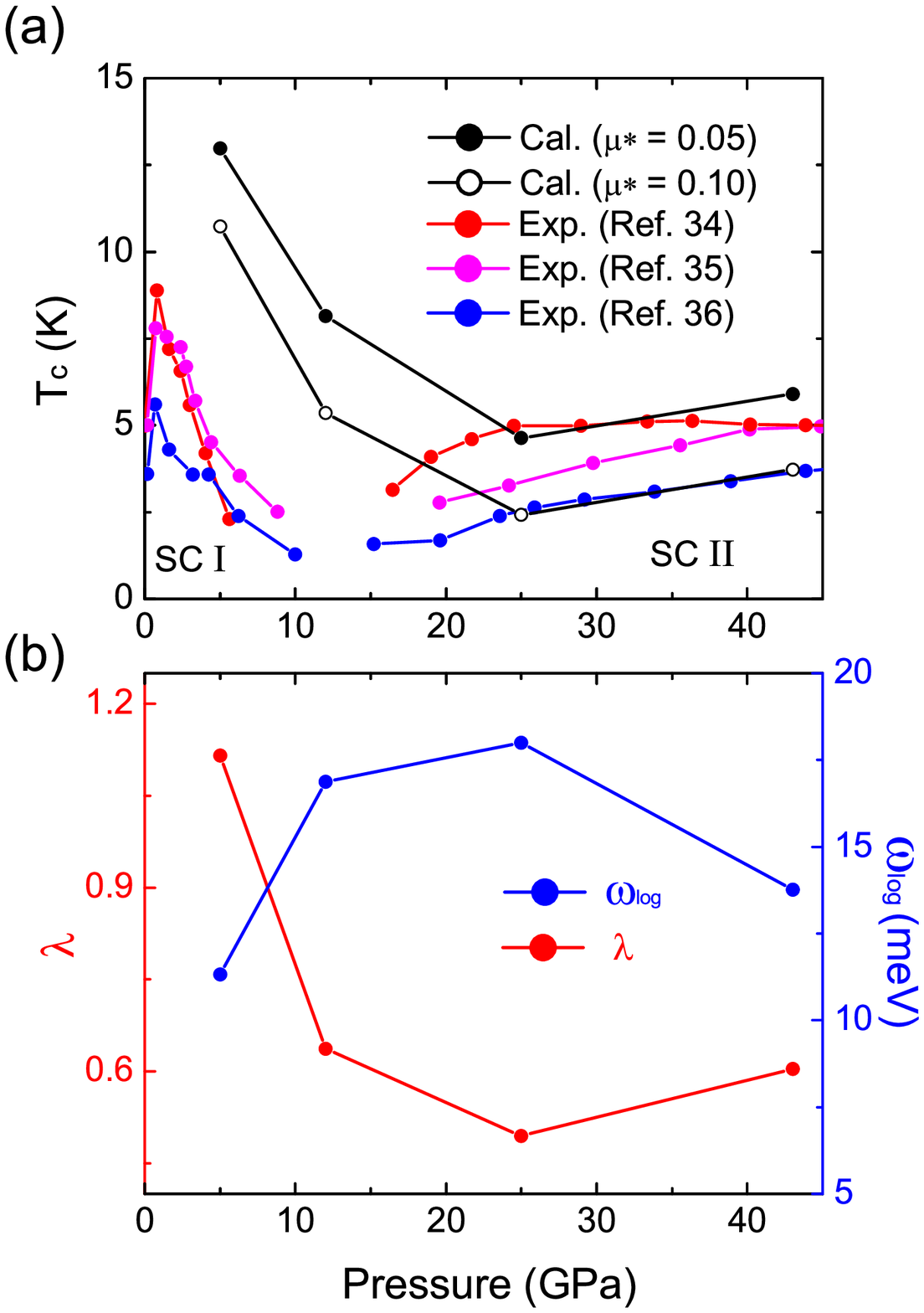}
\caption{(Color online) (a) Superconducting $T_\text{c}$ from our EPC calculations (black dots and circles) and previous experiments~\cite{press1,press2,press3} (red, purple, and blue dots) under pressure. The black dots and circles represent the calculation results when the $\mu^*$ is set to 0.05 and 0.10, respectively. (b) Calculated total EPC constant $\lambda$ and logarithmic frequency $\omega_{\text {log}}$.}
\label{fig3}
\end{figure}

Besides the phonon hardening and softening, our calculations also reveal multiple inversions of the electronic bands after the CDW I state is suppressed ($\sim$ 4 GPa), which changes the topology of the electronic band structure. Specifically , the nontrivial topological invariants Z$_2$ below two curved band gaps around the Fermi level~\cite{z2prl} transform from 1 to 0 at 26 GPa and 40 GPa, respectively. The nontrivial topological band structure before the transition may thus give rise to the possibility of surface topological superconductivity via the proximate effect due to the superconducting bulk. More details about the band inversions and the topological property evolutions under pressure can be found in the Supplemental Materials~\cite{sm}.

Based on the standard EPC calculations and the McMillan-Allen-Dynes formula, we further studied the superconductivity in CsV$_3$Sb$_5$ under the selective pressures without the CDW distortion. The calculated superconducting $T_\text{c}$ at 5, 12, 25, and 43 GPa are shown in Fig. 2(a). The black dots and circles represent the calculation results with the effective Coulomb repulsion $\mu^*$ of 0.05 and 0.10, respectively. The experimental $T_\text{c}$s (red~\cite{press1}, purple~\cite{press2}, and blue~\cite{press3} dots) are also presented for comparison. At 25 and 43 GPa, our calculated results agree well with previous measurements, indicating that the observed superconductivities around these pressures belong to the conventional BCS type. In contrast, the superconducting $T_\text{c}$s derived from the EPC calculations are much higher than the measured ones at 5 and 12 GPa~\cite{press1,press2,press3}, which suggests that some effects are not taken into account in our calculations. We will address this point later. The calculated total EPC strength $\lambda$ and the logarithmic frequency $\omega_{log}$ are shown in Fig. 2(b). Apparently, the pressure evolution of the EPC strength coincides with the behavior of the superconducting $T_\text{c}$. From the moment- and mode- resolved EPC strength shown in Fig. 1, we can learn that the softening modes around the M and L points contribute most to the EPC. The distinct difference between our calculations and previous measurements at low pressures thus needs further exploration.

\begin{figure}[tb]
\includegraphics[angle=0,scale=0.7]{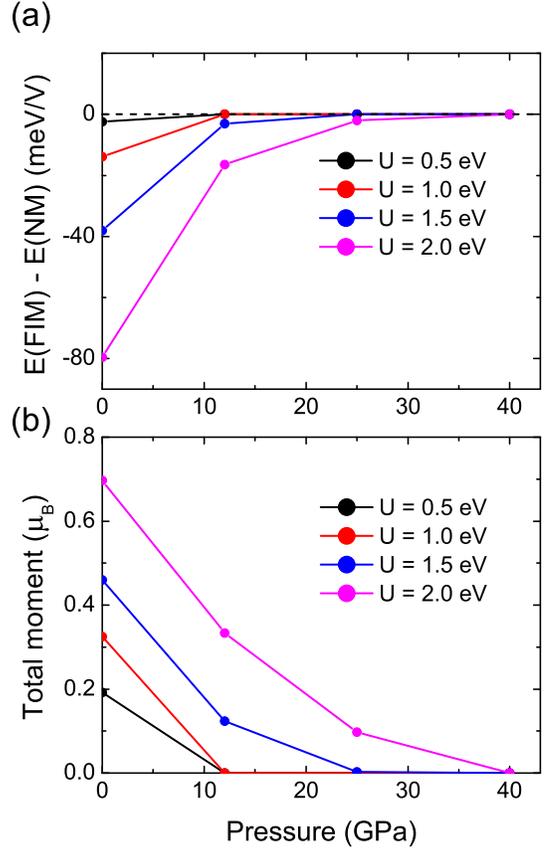}
\caption{(Color online) (a) The pressure dependent energy differences between the ferrimagnetic state and the non-magnetic state ($\Delta E$ = $E(\text{FIM}) - E(\text{NM})$) calculated with different Hubbard U values. (b) The corresponding total magnetic moment in a ferrimagnetic unit cell.}
\label{fig4}
\end{figure}

A natural thought is that the correlations among the V 3$d$ electrons may play some role in the superconductivity at the ambient and low pressures.
Although no static magnetic order has been reported for this family of kagome materials~\cite{1stprm}, there are still some experimental hints for the existence of magnetic interactions, such as the magnetic fluctuations~\cite{mag2,mag} and the orbital order~\cite{mag2}.
We performed the DFT+U calculations to examine the correlation effect and the possible magnetism on the V atoms with some effective Hubbard U values ranging from 0.5 to 2 eV. Several typical collinear or non-collinear magnetic configurations in a 2$\times$2$\times$1 supercell were considered (See Supplemental Materials~\cite{sm} for details). According to our calculations, the relative stability of the magnetic configurations is sensitive to the choice of Hubbard U strengths. With small U values (0.5 and 1.0 eV), a ferrimagnetic state has lower energy than the nonmagnetic and antiferromagnetic states. Figure 3(a) shows the pressure evolution of the energy difference between the ferrimagnetic state and the nonmagnetic state [$\Delta E=E$(FIM)$-$$E$(NM)]. The corresponding total magnetic moments in the unit cell are between 0.2 $\mu_B$ and 0.7 $\mu_B$ at ambient pressure, as shown in Fig. 3(b). According to the BCS theory, the existence of the ferrimagnetism or ferrimagnetic fluctuations is harmful to the conventional superconductivity. This may explain the lower observed $T_\text{c}$s than our calculated values at low pressures [Fig. 2(a)]. As the pressure increases, the magnetism on the V atoms can be effectively suppressed [Fig. 3(b)], driving the system to a nonmagnetic state and meanwhile reviving the conventional superconductivity [Fig. 2(a)].

\begin{figure}[tb]
\includegraphics[angle=0,scale=0.36]{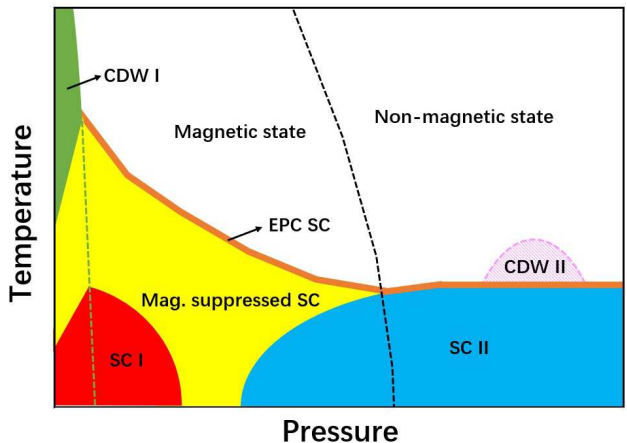}
\caption{(Color online) A schematic superconducting phase diagram of CsV$_3$Sb$_5$. The green region represents the original CDW state. The red and blue regions label the experimentally observed double superconducting domes respectively. The orange line represents the conventional EPC-derived superconducting $T_\text{c}$ without the magnetism. The black dashed line outlines the pressure-suppressed magnetic state. The yellow region represents the magnetism-suppressed superconductivity. The pink shaded region covers our predicted second CDW dome.}
\label{fig5}
\end{figure}

We simplify our opinion by a schematic superconducting phase diagram in Fig. 4. The green region represents the CDW state (CDW I) at ambient pressure. The red and blue regions show the experimentally observed double superconducting domes. The orange line represents our expected conventional EPC superconducting $T_\text{c}$ without any magnetism and the black dashed line outlines the pressure-suppressed magnetic state. At ambient/low pressure, the CDW I state can repulse the electronic states around the Fermi level and partly suppress the superconductivity~\cite{yan}, which is consistent with the competitive relation between the CDW state and the superconducting state in many CDW materials~\cite{TaS2,TaSe2,TiSe2}. In addition to the CDW state, the hidden ferrimagnetism or other possible magnetic phases are also harmful to the conventional superconductivity. As a result, the observed superconducting $T_\text{c}$s are far below our theoretical expectation (represented by yellow region) and even completely vanishes with the pressure, forming the first superconducting dome. In fact, the experimental pressure at which the CDW is suppressed (0.8 GPa)~\cite{press1} is smaller than our calculated one [4 GPa in Fig. 1(f)], which may come from the restrained effect of the magnetism on the CDW state~\cite{monoNbSe2,VS2}. It's also worth noting that the experimentally proposed multiple-type (nodal or nodeless) superconducting gaps~\cite{lowpress2,sc2,mag,fengstm} may likely be due to the coexistence of the ferrimagnetic fluctuations and the $s$-wave superconductivities. At high pressure, the magnetism is completely suppressed and the system restores its original conventional superconductivity, forming the second superconducting dome. Interestingly, our nonmagnetic calculation at high pressure also predicts a transitory weak CDW state (CDW II). Nevertheless, its distortion is distinctly different from the CDW I state, which needs future experimental verification.

\section{Summary}

In summary, we have studied the pressure evolution of the CDW state, superconductivity, and topological properties of the kagome material CsV$_3$Sb$_5$ by using first-principles electronic structure calculations. Our calculations show that the CDW distortion in CsV$_3$Sb$_5$ is rapidly suppressed at low pressure. The calculated superconducting $T_\text{c}$ based on the conventional EPC theory demonstrates an interesting behavior: it decreases gradually from 5 to 25 GPa (low pressure range) and stays stable between 25 and 43 GPa (high pressure range). Our calculated $T_\text{c}$'s in the high pressure range are in good accordance with the experimental observations, however, the low-pressure values show distinct difference~\cite{press1,press2,press3}. By performing the DFT+U calculations, we find that the modest electron-election correlation can induce the magnetism on the V atom and the pressure can effectively suppress it. Based on these results, we deduce that the experimentally observed superconductivity in CsV$_3$Sb$_5$ at ambient/low pressure may belong to the conventional BCS type but are partially suppressed by the magnetism (very likely ferrimagnetism) on the V atoms, while those at high pressure are fully of the BCS type with vanishing local moments. In addition, we also predict a weak 2$\times$2$\times$1 CDW dome and the topological phase transition from nontrivial to trivial states at high pressures. Our studies may apply to other kagome materials KV$_3$Sb$_5$ and RbV$_3$Sb$_5$, which show similar superconducting behaviors~\cite{press3}, and wait for further experimental confirmation.

\begin{acknowledgments}

This work was supported by the National Key R\&D Program of China (Grants No. 2017YFA0302903 and No. 2019YFA0308603), the Beijing Natural Science Foundation (Grant No. Z200005), the National Natural Science Foundation of China (Grants No. 11774424 and No. 11934020), the CAS Interdisciplinary Innovation Team, the Fundamental Research Funds for the Central Universities, and the Research Funds of Renmin University of China (Grant No. 19XNLG13). Computational resources were provided by the Physical Laboratory of High Performance Computing at Renmin University of China.

\end{acknowledgments}

\end{document}